\documentclass[10pt, conference, letterpaper]{IEEEtran}
\ifCLASSINFOpdf
\else
\fi
\usepackage[numbers,sort&compress]{natbib}
\usepackage{amsfonts}
\usepackage{amsmath}
\usepackage{amssymb}
\usepackage{subeqnarray}
\usepackage{cases}
\usepackage{array}
\usepackage[ruled,linesnumbered]{algorithm2e}
\usepackage{algorithmic}
\usepackage{bm}
\usepackage{cuted}
\usepackage{multicol}
\usepackage[numbers,sort&compress]{natbib}
\usepackage{extarrows}

\usepackage{multirow}
\usepackage{makecell,diagbox}
\usepackage{balance}
\usepackage{graphicx}
\usepackage{subfigure}
\usepackage{mathrsfs}
\usepackage{ntheorem}
\usepackage{url}
\graphicspath{{pic/}}

\theoremheaderfont{\itshape}
\theorembodyfont{\normalfont\rm}

\newtheorem{definition1}{\hspace{2em}Definition}
\theoremseparator{:}

\newcommand{\PreserveBackslash}[1]{\let\temp=\\#1\let\\=\temp}
\newcolumntype{C}[1]{>{\PreserveBackslash\centering}p{#1}}
\newcolumntype{R}[1]{>{\PreserveBackslash\raggedleft}p{#1}}
\newcolumntype{L}[1]{>{\PreserveBackslash\raggedright}p{#1}}

\begin{document}
%
\title{Enabling Quality-Driven Scalable Video Transmission over Multi-User NOMA System}
%
%
%

\author{\IEEEauthorblockN{Xiaoda Jiang$^{\dag}$, Hancheng Lu$^{\dag}$, Chang Wen Chen$^{\ddag}$$^{\S}$}
\IEEEauthorblockA{$^{\dag}$Key Laboratory of Wireless-Optical Communications, Chinese Academy of Sciences,\\
School of Information Science and Technology, University of Science and Technology of China, China\\
$^{\ddag}$School of Science and Engineering, Chinese University of Hong Kong, Shenzhen, China\\
$^{\S}$Department of Computer Science and Engineering, State University of New York at Buffalo, USA\\
jxd95123@mail.ustc.edu.cn, hclu@ustc.edu.cn, chencw@cuhk.edu.cn
}}
\maketitle
\begin{abstract}
Recently, non-orthogonal multiple access (NOMA) has been proposed to achieve higher spectral efficiency over conventional orthogonal multiple access. Although it has the potential to meet increasing demands of video services, it is still challenging to provide high performance video streaming. In this research, we investigate, for the first time, a multi-user NOMA system design for video transmission. Various NOMA systems have been proposed for data transmission in terms of throughput or reliability. However, the perceived quality, or the quality-of-experience of users, is more critical for video transmission. Based on this observation, we design a quality-driven scalable video transmission framework with cross-layer support for multi-user NOMA. To enable low complexity multi-user NOMA operations, a novel user grouping strategy is proposed. The key features in the proposed framework include the integration of the quality model for encoded video with the physical layer model for NOMA transmission, and the formulation of multi-user NOMA-based video transmission as a quality-driven power allocation problem. As the problem is non-concave, a global optimal algorithm based on the hidden monotonic property and a suboptimal algorithm with polynomial time complexity are developed. Simulation results show that the proposed multi-user NOMA system outperforms existing schemes in various video delivery scenarios.
\end{abstract}


%
\IEEEpeerreviewmaketitle

\textfloatsep=10pt plus 0pt minus 0pt

\begin{figure*}[htb]
\centering
\includegraphics[width=12cm]{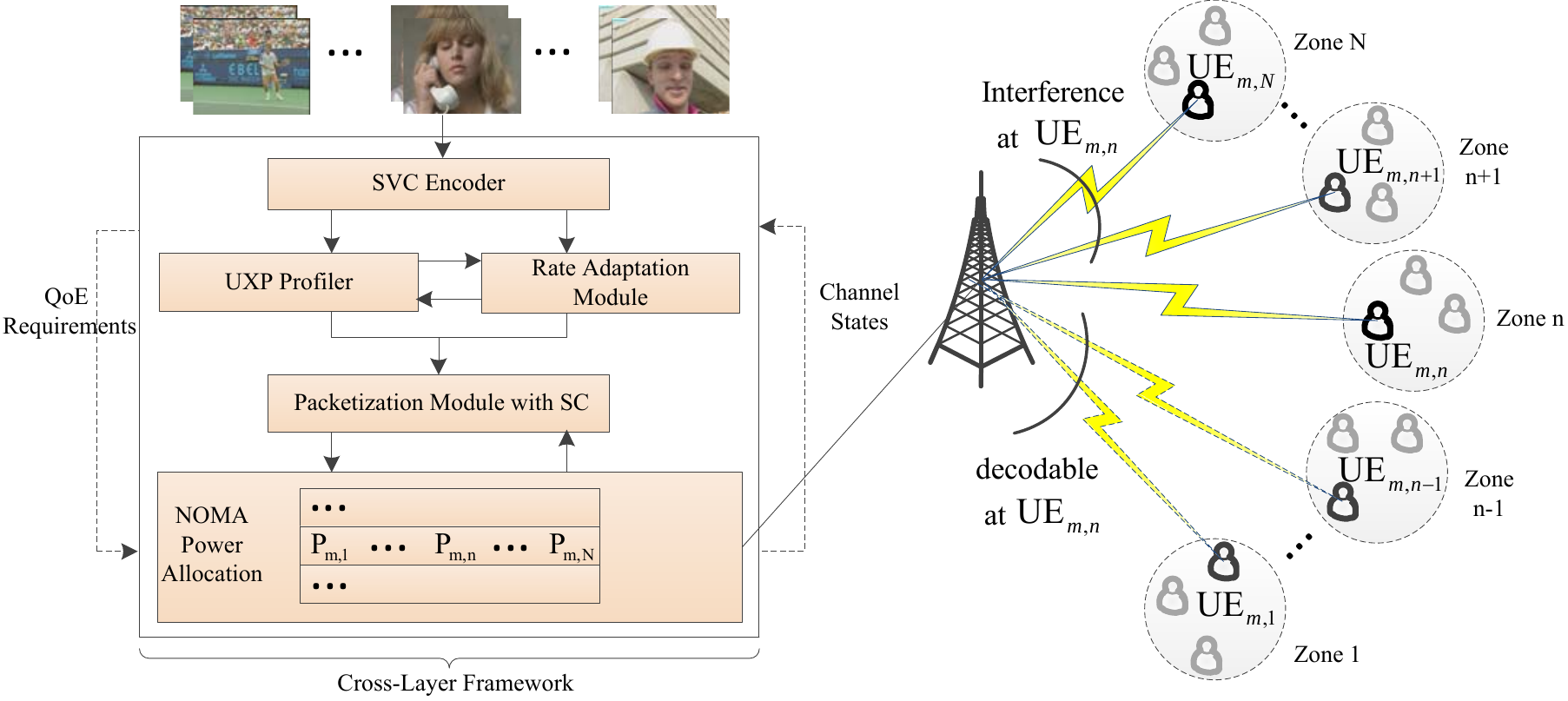}
\vspace{-0.2cm}
\caption{Architecture of  multi-user video transmission over NOMA system.}\label{figure:systemOverview}
\vspace{-1em}
\end{figure*}

\section{Introduction}
%
%
%
%
Recently, non-orthogonal multiple access (NOMA) has emerged as a promising technology in 5th generation (5G) mobile systems, to meet the requirements of ultra-high capacity and ultra-low latency in 5G \cite{white2014rethink}, \cite{Higuchi2015non}, \cite{islam2017power}. Different from orthogonal multiple access (OMA), NOMA can serve users with the same channel resources (e.g., time and frequency), achieving significant spectral efficiency improvements. These advantages are obtained as NOMA explores the power domain for multiple access. Specifically, at the sender, signals of NOMA users are multiplexed by the superposition coding (SC) with different power. Meanwhile, at the receiver, successive interference cancellation (SIC) \cite{caire2003on} is performed to remove the undesired multi-user interference (MUI). With SIC, better users with stronger channel gains can help decode information of weaker users with worse channel gains.
\par
Many studies have been carried out to analyze and improve performance of NOMA systems \cite{Higuchi2015non}, \cite{islam2017power}, \cite{wang2016power}, \cite{ding2016impact}, \cite{lv2016cooperative}, \cite{elbamby}. The authors in \cite{Higuchi2015non} showed that NOMA can outperform OMA in terms of spectral efficiency. A power allocation algorithm was proposed in \cite{wang2016power} to maximize sum throughput with the quality-of-service (QoS) requirement of the weaker user. Performance of NOMA was investigated with cooperative transmission in \cite{lv2016cooperative}. Furthermore, the authors in \cite{elbamby} investigated power allocation of uplink and downlink jointly in full duplex NOMA. However, early work mainly focused on NOMA networks with data transmission. There is a lack of work in considering true video applications in NOMA. In contrast, video traffic will contribute over 75\% of total mobile traffic in 2020 \cite{inc2016cisco}.
\par
Unfortunately, the design principle of existing NOMA systems for data transmission cannot be easily adopted for video applications, because there exist two critical challenges in applying NOMA in video transmission. First, in most existing studies, NOMA optimization is performed only based on channel conditions of the physical (PHY) layer. However, it is well known that the video content at the application (APP) layer could significantly influence the quality-of-experience (QoE) of the users. Due to the content-dependency of video streams, different bits may own different importance to video quality, which calls for unequal erasure protection (UXP) \cite{schierl2005wireless} at the APP layer. Therefore, a cross-layer design in NOMA systems is necessary for video delivery, with the consideration of both content characteristics at the APP layer and channel conditions at the low layers. Second, power allocation in existing NOMA studies aims at optimizing network performance (e.g., throughput, reliability). However, for video transmission, the perceived video quality from the user perspective, instead of common network performance, has been highlighted for characterizing the satisfaction of video delivery. Moreover, diverse user equipments (UEs) impose various quality requirements of video streams. Therefore, quality-driven power allocation is essential for video transmission in NOMA.
\par

To overcome aforementioned challenges, in this research, we consider quality-driven optimization on the multi-user NOMA system for video transmission. To our best knowledge, this is the first attempt to adopt NOMA for true video delivery. The main contributions are summarized as follows.
\par
1) A quality-driven cross-layer framework, with consideration of the characteristics of encoded video sequences as well as the SC in the NOMA system, is developed for scalable video transmission. To reduce complexity of NOMA implementation in the multi-user scenario, a user grouping strategy is proposed, based on user locations, perceived quality requirements, and requested video contents of UEs. By doing so, intense MUI can be alleviated, and the latency resulting from SIC can also be reduced.
\par
2) In the proposed framework, we integrate a semi-analytical quality model for scalable video streams measured in terms of peak signal-to-noise ratio (PSNR), with the PHY model in the multi-user NOMA system. Based on this, the quality-driven power allocation problem for the multi-user NOMA system is formulated as an average PSNR maximization problem under diverse quality constraints from users.
\par
3) The quality-driven power allocation problem has been shown to be non-concave. To solve it tractably, we explore the hidden monotonic property and design an efficient algorithm to find the global optimal solution. Furthermore, we also design a suboptimal algorithm, inspired by the process of SIC and greedy strategy, to approach the solution with polynomial time.
\par
Extensive simulations have been carried out and the results validate the advantage of the proposed multi-user NOMA system. It can achieve significant gains for scalable video transmission in terms of PSNR, compared with conventional video delivery schemes over OMA networks as well as existing NOMA schemes developed for data transmission. Moreover, simulation results also provide guidance for designing dynamic user grouping strategies in mobile scenarios, based on characteristics of requested contents.
\par
The rest of the paper is organized as follows. In Section \ref{section:System Description}, we describe the system architecture for quality-driven cross-layer scalable video transmission over the NOMA system. In Section \ref{section:Problem Formulation and Analysis}, the quality-driven power allocation problem is formulated, based on the quality model using PSNR as a metric and the PHY model in NOMA. Both optimal algorithm and suboptimal algorithm with polynomial time are described in Section \ref{section:solution}. Performance evaluations are presented in Section \ref{section:performance-evaluation}. In Section \ref{section:conclusion}, we conclude this paper with a summary.
\par
{\em Notations:} Bold and calligraphic letters denote vectors and sets, respectively. $\bf 0$ represents the all-zero vector. $\mathcal{R}^n$ and $\mathcal{R}_+^n$ denote the set of $n$-dimensional real and nonnegative real vectors, respectively. Given vectors ${\bf x},{\bf y} \in \mathcal{R}^n$, ${\bf x}\preceq{\bf y}$ implies that $x_i\leq y_i$, for all $1\leq i \leq n$. $\cap$, $\cup$ and $\backslash$ represent set intersection, set union and set difference operators, respectively.

\section{System Description}\label{section:System Description}
In this section, we present the architecture of the proposed quality-driven scalable video transmission framework over the multi-user NOMA system, as shown in Fig. \ref{figure:systemOverview}. We shall also elaborate the cross-layer design adapted to the SC in NOMA.
\par
To satisfy different quality requirements of diverse sets of UEs, scalable video coding (SVC) \cite{schwarz2007overview} is adopted in the proposed system, providing scalable substreams for the SC. To illustrate compactly such scalability, we adopt quality scalable video streams, with the fixed frame rate and resolution. Two ways of quality scalability are available in SVC standards and implementations \cite{JSVM}, namely coarse-grain quality scalable coding (CGS) and medium-grain quality scalable coding (MGS). CGS provides desired scalability by successively dropping quality layers until the target bit rate is met. However, it only provides a number of choices limited to the number of CGS quality layers. In this research, we adopt MGS, which allows finer scalability by dividing each CGS layer into up to 16 MGS layers.
\par
The basic unit for video coding is a group of pictures (GOP). In this research, we assume that each GOP begins with an intra-coded frame (I-frame), which ensures the intra-refresh period is the same as the GOP size $G$. Therefore, throughout the paper, the processing of video sequences and optimization of power allocation are carried out with the GOP interval.
\subsection{Multi-User NOMA System with User Grouping}\label{subsection:NOMA Wireless Network with User Grouping}
We consider a downlink single-cell communication scenario with one base station (BS) and multiple UEs of various quality requirements. As NOMA is MUI limited, it is not realistic to perform NOMA among all UEs simultaneously \cite{ding2016impact}. Especially for delay-sensitive video services, too many superimposed streams would impose serious computational complexity for SIC. The increase in decoding time may degrade the QoE of UEs. We believe that it is beneficial to design a hybrid multiple access scheme in which NOMA is combined with OMA. Specifically, we first divide UEs into $N$ zones, each of which is equally deployed with $M$ UEs. Prior to transmission, the BS selects one UE from each zone and groups these $N$ selected UEs together. This would result in $M$ groups. Then NOMA is implemented in each group, while orthogonal bandwidth resources are allocated among groups, as shown in Fig. \ref{figure:systemOverview}.
\par
The spirit of grouping multiple users for NOMA presented above may be traced in an early study in \cite{ding2016impact}. However, the study in \cite{ding2016impact} deals with only two users paired for NOMA. Moreover, the critical difference between the proposed strategy and the study in \cite{ding2016impact} is that the proposed scheme is based on not only locations, but also received quality requirements and requested contents of UEs. UEs with distinct locations are first divided into $N$ zones, according to their connection quality to the BS, i.e., the $n^{th}$ zone has better connection quality to the BS than the $n-1^{th}$ zone. Then for the UE located at the edge of two adjacent zones, the selection of its subordinate zone is decided based on its quality requirement. For example, a UE is close to both $n^{th}$ and $n-1^{th}$ zones, and playback delay is an important factor on its perceptive quality. Consider that a user in the $n^{th}$ zone need to detect information of $n-1$ users via SIC, while a user in the $n-1^{th}$ zone only detects information of $n-2$ users via SIC. Thus, this UE should be partitioned into the $n-1^{th}$ zone, with lower SIC latency. Moreover, for mobile scenarios, our proposed strategy dynamically groups weaker UEs requesting sequences of lower spatial-temporal content complexity, together with better UEs requesting sequences of higher content complexity. The advantages of such content-based grouping strategy are confirmed in \ref{subsection:impacts of user grouping}.
\par
The $m^{th}$ NOMA group $\{{\rm{UE}}_{m,1},{\rm{UE}}_{m,2},\cdots,{\rm{UE}}_{m,N}\}$ is considered as an example in Fig. \ref{figure:systemOverview}. As we have assumed, the UEs' channel quality in this group is ordered as $|h_{m,1}|^2\leq|h_{m,2}|^2\leq\cdots\leq|h_{m,N}|^2$, where $h_{m,n}$ is the fading channel gain between the BS and the $n^{th}$ UE in the $m^{th}$ NOMA group. In particular, it is assumed that $h_{m,n}=g_{m,n}/\sqrt{1+d_{m,n}^{\eta}} $, where $d_{m,n}$ is the distance from ${\rm{UE}}_{m,n}$ to the BS, $\eta$ is the path-loss exponent and $g_{m,n}\sim\mathcal{CN}(0,1)$ is the Rayleigh fading coefficient.
\par
Based on the proposed user grouping strategy, we now discuss the procedure for signal decoding at ${\rm{UE}}_{m,n}$. With SIC, the superimposed signals of $\{{\rm{UE}}_{m,1},\cdots,{\rm{UE}}_{m,n-1}\}$ are decoded and subtracted before ${\rm{UE}}_{m,n}$ decodes its own signal. The superimposed signals of $\{{\rm{UE}}_{m,n+1},\cdots,{\rm{UE}}_{m,N}\}$ are treated as interference noise. Therefore, the received signal-to-interference-plus-noise ratio (SINR) at ${\rm{UE}}_{m,n}$ to detect the signal of ${\rm{UE}}_{m,\tilde{n}}$ $(\tilde{n}\leq n)$ will be:
\begin{equation}\label{equation:SINR}
\setlength{\abovedisplayskip}{4pt}
\setlength{\belowdisplayskip}{4pt}
\gamma_{m,n}^{\tilde{n}}\!({\bf P}\!_m)\!=\!
\begin{cases}
    \frac{|h_{m,n}|^2P_{m,\tilde{n}}}{\sigma^2},&\tilde{n}=n=N,\\
    \frac{|h_{m,n}|^2P_{m,\tilde{n}}}{\sum_{i=\tilde{n}+1}^{N}|h_{m,n}|^2P_{m,i}+\sigma^2}, &1\leq \tilde{n}\leq n \leq N-1,
\end{cases}
\end{equation}
where ${\bf P}\!_m=(P_{m,1},\cdots,P_{m,N})$ is the allocated power vector in the $m^{th}$ NOMA group. $\sigma^2$ is the variance of the additive white Gaussian noise (AWGN), which is assumed to be the same among all UEs. Throughout the paper, channel state information (CSI) is assumed to be perfectly known.

\subsection{Quality-Driven Cross-Layer Operations with SC}\label{subsection:QoE-based Cross-Layer Structure with SC}
As Fig. \ref{figure:systemOverview} illustrates, quality-driven power allocation is performed at the lower layers, with interactive information from higher layers. At the APP layer, each video sequence is first encapsulated into network abstraction layer units (NALUs). UXP is then carried out on these NALUs, to achieve robustness over packet lossy networks. This process is based on content characteristics and the real-time transport protocol (RTP) packet loss rate $p_{rtp}$. Knowledge of $p_{rtp}$ can be obtained via short term estimation based on CSI. At the rate allocation module, the unnecessary NALUs of each stream are dropped, based on the estimated channel capacity after quality-driven power allocation. The remaining NALUs are organized to form a certain substream for the SC.


\begin{figure}[htb]
\vspace{-0.1cm}
\centering
\includegraphics[width=6.1cm]{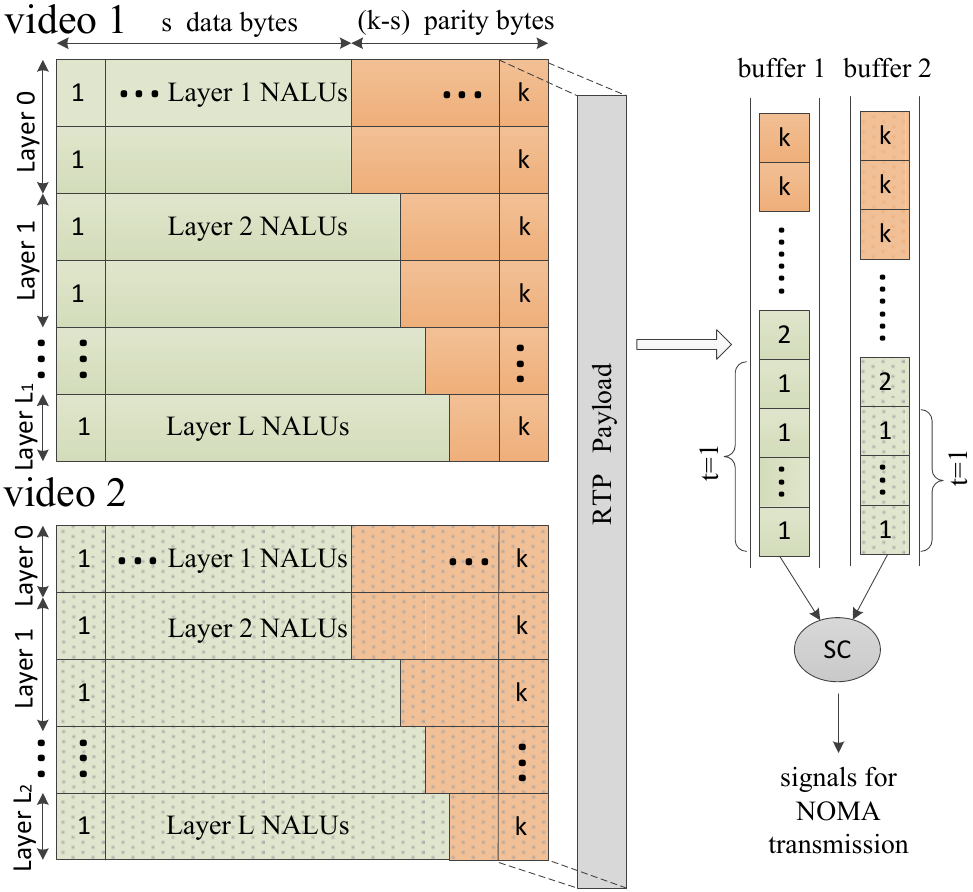}
\vspace{-0.2cm}
\caption{TSB structure, RTP packetization and SC for two video streams to two NOMA UEs.}\label{figure:TSB-structure}
\vspace{-0.1cm}
\end{figure}
\par
Due to the SC in the NOMA system, the packetization should be carefully reconsidered. Without loss of generality, the proposed design is illustrated for the case of two NOMA UEs. Each GOP after the UXP process is mapped into one transmission subblock (TSB) with data and parity bytes. The row of a TSB is a ($k,s$) Reed-Solomon (RS) codeword. As streams of the two NOMA UEs are superposed, we select one TSB from each encoded stream requested by its corresponding UE. Then we encapsulate these two TSBs into a transmission block (TB), whose column is the payload of a RTP packet. This is illustrated in Fig. \ref{figure:TSB-structure}. Note that the division of zones should be designed carefully, to ensure that the RTP payload does not exceed the size supported by the network protocol.
\par
Bytes in TSBs of the two UEs are stored by column order in buffer 1 and 2, respectively. Starting with the first timeslot, bytes belonging to the first column of TSB in buffer 1, together with corresponding bytes in buffer 2, are modulated and superposed to form the NOMA signals, as illustrated in Fig. \ref{figure:TSB-structure}. The same superposition process will be repeated for the bytes in two buffers, belonging to the $2^{th}$ column of TSBs, in the following available timeslot, and so forth.

\section{Problem Formulation and Analysis}\label{section:Problem Formulation and Analysis}
To design quality-driven optimization in the multi-user NOMA system for scalable video transmission, we need to first investigate the quality model for video transmission and the PHY layer model for the NOMA system. We will then formulate the quality-driven power allocation problem in order to maximize the average perceived quality of all users.

\subsection{Quality Model for Encoded Video Streams}\label{subsection:QoE Model for Encoded Video Streams}
For video applications, PSNR is an objective parameter, which has been shown highly correlated with user-perceived video quality \cite{vqeg2003final}. In this research, we shall adopt PSNR as the metric that measures received video quality.
\par
The relation between PSNR and distortion (i.e., mean squared error - MSE) is defined as:
\begin{equation}\label{equation:PSNR}
{\rm PSNR=10\log_{10}(255^2/{MSE})}.
\end{equation}
Thus, we need first analyze the rate-distortion (R-D) model. A general continuous semi-analytical R-D model was proposed in \cite{stuhlmuller2000analysis} to estimate the relationship between the rate and distortion at the video encoder. This operational relationship has been verified in \cite{mansour2008channel} for SVC MGS layers. After converting distortion to PSNR, the rate of the video stream for ${\rm{UE}}_{m,n}$ as the function of PSNR will be:
\begin{equation}\label{equation:curve-fitting}
F_{m,n}(Q)=\frac{\theta_{m,n}}{255^210^{(-Q/10)}+\alpha_{m,n}}+\beta_{m,n},
\end{equation}
where $Q$ is the PSNR value. The parameters $\alpha_{m,n}$, $\beta_{m,n}$ and $\theta_{m,n}$ are dependent on the video content, specific video encoder and RTP packet loss rate. And they can be estimated with curve-fitting over at least six empirical R-D points \cite{mansour2008channel}.
\par
Note that the video streams are processed with UXP for packet lossy networks. Thus, the rate should be calculated with parity bytes. While distortion $D$ includes encoder distortion $D_{enc}$ and transmission distortion $D_{tran}$. Because $D_{enc}$ and $D_{tran}$ can be considered uncorrelated, end-to-end distortion can be estimated at the encoder as: $D=D_{enc}+D_{tran}$ \cite{he2002joint}. Therefore, the R-D point can be estimated at the APP layer, and subsequently the relationship between PSNR and the rate can be established according to (\ref{equation:PSNR}) and (\ref{equation:curve-fitting}).

\subsection{Physical Layer Model for NOMA System}\label{subsection:Physical Layer Model for NOMA wireless networks}
As discussed in \ref{subsection:NOMA Wireless Network with User Grouping}, orthogonal bandwidth resources are allocated among NOMA groups, and all UEs in a group are implemented with NOMA. Assume that the bandwidth assigned to $m^{th}$ group, corresponding to a GOP transmission period, is $B_m$. Then, ${\rm{UE}}_{m,1},\cdots,{\rm{UE}}_{m,N}$ will share the $B_m$ downlink bandwidth with power budget $P^{max}_m$ for transmitting their superposed NOMA signals.
\par
With SIC, the SINR $\gamma_{m,n}^{\tilde{n}}\!({\bf P}\!_m)$, for ${\rm{UE}}_{m,n}$ detecting the signal of ${\rm{UE}}_{m,\tilde{n}}$ $(\tilde{n}\leq n)$, is given by (\ref{equation:SINR}). We assume that $\gamma_{m,n}^{\tilde{n}}\!({\bf P}\!_m)$ remains constant within one GOP duration, but varies across GOP durations. Considering the adaptive modulation and coding (AMC) scheme adopted in the PHY layer, the achievable rate for ${\rm{UE}}_{m,n}$ to detect ${\rm{UE}}_{m,\tilde{n}}$ $(\tilde{n}\leq n)$ is given by:
\begin{equation}\label{equation:PHY-rate}
R_{m,n}^{\tilde{n}}({\bf P}_m)=c_1 B_m \log(1+\gamma_{m,n}^{\tilde{n}}({\bf P}_m)/c_2),
\end{equation}
where $c_1$ and $c_2$ are {\em rate adjustment} and {\em SNR gap}, respectively, determined by a practical AMC scheme \cite{mazzotti2012multiuser}.
\par
Without loss of generality, we can assume that the overhead introduced by each network layer is constant and can be ignored. In this case, the PHY rate will be equal to the APP layer rate, i.e.,
$R_{m,n}^{{n}}({\bf P}_m)=F_{m,n}(Q)$. According to (\ref{equation:curve-fitting}) and (\ref{equation:PHY-rate}), the relationship between PSNR of the stream delivered to ${\rm UE}_{m,n}$ and its corresponding PHY rate, can be written as:
\begin{equation}\label{equation:PSNR-PHY-rate}
\setlength{\abovedisplayskip}{3pt}
\begin{aligned}
\!Q_{m,n}\!=&F_{m,n}^{-1}(R_{m,n}^{{n}}({\bf P}_m))\\
=&\!-\!10\log_{10} (\frac{\theta_{m,n}}{c_1\! B_m \! \log(1\!\! +\! \!\gamma_{m,n}^{n}({\bf P}_m)/c_2)\! \!-\! \!\beta_{m,n}}\!-\!\alpha_{m,n})\\
&\! +\! 20\log_{10}255.
\end{aligned}
\end{equation}
\vspace{-0.6cm}
\subsection{Quality-Driven Power Allocation Problem}\label{subsection:Distortion-Aware Power Allocation Problem for Video transmission in NOMA wireless networks}
Note that orthogonal bandwidth resources are assumed equally allocated among NOMA groups. Therefore, we shall focus on the power allocation problem within each NOMA group, e.g., the $m^{th}$ group. In the subsequent analysis, we omit the group index $m$ for brevity.
\par
Based on (\ref{equation:PSNR-PHY-rate}), the quality-driven power allocation problem in the NOMA system can be formulated as a constrained PSNR maximization problem:
\begin{subequations}\label{equation:original-optimization}
\setlength{\abovedisplayskip}{-1pt}
\setlength{\belowdisplayskip}{6pt}
\begin{align}
\max_{{\bf P}} ~~ &Q({\bf P}) = \frac{1}{N}\sum_{n=1}^{N}Q_n({\bf P})\label{equation:original-optimization-a}\\
\text{s.t.}~~~\ &\sum_{n=1}^{N}  P_{n} \leq P^{max},\label{equation:original-optimization-b} \\
&0 \leq P_{n}, ~ \forall n \in \mathcal{N},\label{equation:original-optimization-c}\\
&Q_{n}^{min} \leq Q_{n}({\bf P}) \leq Q_{n}^{max}, ~ \forall n \in \mathcal{N},\label{equation:original-optimization-d}\\
&\gamma_{\tilde{n}}^{min} \leq \gamma_{n}^{\tilde{n}}({\bf P}), ~ \forall \tilde{n}, n \in \mathcal{N}, ~\tilde{n} < n,\label{equation:original-optimization-e}
\end{align}
\end{subequations}
where $Q_n({\bf P})$ is expressed in (\ref{equation:PSNR-PHY-rate}). $\mathcal{N}$ is the index set of NOMA UEs. (\ref{equation:original-optimization-b}) and (\ref{equation:original-optimization-c}) are system power constraints. In (\ref{equation:original-optimization-d}), the maximum PSNR $Q_n^{max}$ and the minimum PSNR $Q_n^{min}$ depend on the video content, encoder and the quality requirement of ${\rm UE}_n$. (\ref{equation:original-optimization-e}) is the SIC constraint in NOMA, which guarantees that signals for weaker UEs are still decodable at better UEs.
\par
According to (\ref{equation:curve-fitting}) and (\ref{equation:PHY-rate}), the PSNR function is strictly monotone in terms of the SINR. Thus the quality requirements ($Q_n^{min},Q_n^{max}$) can be replaced by corresponding SINR constraints ($\gamma_{n}^{min},\gamma_{n}^{max}$). Since $|h_{\tilde{n}}|^2 \leq |h_n|^2$ for $\tilde{n} < n$ as assumed in \ref{subsection:NOMA Wireless Network with User Grouping}, according to (\ref{equation:SINR}), (\ref{equation:original-optimization-e}) will be satisfied as long as (\ref{equation:original-optimization-d}) is satisfied. Hence, (\ref{equation:original-optimization-e}) can be removed.
\par
To simplify expression, we rewrite the SNIR, for the $n^{th}$ UE to detect its own video stream, as $\gamma_{n}({\bf P})$. The functions $f_{n}({\bf x}),\xi_{n}({\bf x}): \mathcal{R}^N \rightarrow \mathcal{R}_{++}$ can be defined as follows:
\begin{subequations}\label{equation:fraction-function}
\begin{align}
f_{n}({\bf x}) &= |h_n|^2x_n,\\
\xi_{n}({\bf x}) &=
\begin{cases}
    \sigma^2,&n=N,\\
    \sum_{i=n+1}^{N}|h_n|^2x_i+\sigma^2, &1\leq n \leq N-1.
\end{cases}
\end{align}
\end{subequations}
Then the problem in (\ref{equation:original-optimization}) is equivalent to the following problem:
\begin{subequations}\label{equation:GLFP-optimization}
\setlength{\abovedisplayskip}{-1pt}
\setlength{\belowdisplayskip}{4pt}
\begin{align}
\max_{{\bf P}} \quad &\psi(\frac{f_1({\bf P})}{\xi_1({\bf P})},\cdots,\frac{f_N({\bf P})}{\xi_N({\bf P})})\label{equation:GLFP-optimization(a)}\\
\text{s.t.}~\quad & \sum_{n=1}^{N} P_{n} \leq P^{max},\label{equation:GLFP-optimization(b)}\\
&0 \leq P_{n}, \; \forall n \in \mathcal{N},\label{equation:GLFP-optimization(c)}\\
&\gamma_{n}^{min} \leq \gamma_{n}({\bf P}) \leq \gamma_{n}^{max}, \; \forall n \in \mathcal{N},\label{equation:GLFP-optimization(d)}
\end{align}
\end{subequations}
where function $\psi({\bf x})$ is increasing on $\mathcal{R}_+^N$, expressed as:
\begin{equation}\label{equation:phi-function}
\psi({\bf x})=-\frac{10}{N}\log_{10}\prod_{n=1}^{N}  (\frac{\theta_{n}}{c_1 B  \log(1+ x_n/c_2)-\beta_{n}}-\alpha_{n})
\end{equation}

\par
The problem in (\ref{equation:GLFP-optimization}) actually belongs to the class of the General Linear Fractional Programming (GLFP) problem. Due to the characteristics of functions $f_{n}({\bf x}),\xi_{n}({\bf x})$ and $\psi({\bf x})$, this problem is a non-concave optimization problem. In general, there is no efficient method to find the global optimal solution. However, in the next section, we will show how to exploit the hidden monotonicity of the problem in (\ref{equation:GLFP-optimization}), to design a tractable quality-driven optimal power allocation algorithm.

\section{Solutions to the Quality-Driven Power Allocation Problem}\label{section:solution}
In this section, we propose a global optimal algorithm for the problem in (\ref{equation:GLFP-optimization}) by exploiting the monotonic optimization theory to remedy computational complexity. Furthermore, we design a suboptimal algorithm which can approach the optimal solution with polynomial time.

\subsection{Global optimal solution}
\subsubsection{Monotonic Optimization}
First, we will introduce some mathematical definitions that will be useful in monotonic optimization \cite{tuy2000Monotonic}, \cite{zhang2013Monotonic}.

\begin{definition1}\label{def:box}
\emph{(Box):} Given any vector ${\rm{\bf z}} \in \mathcal{R}^N_+$, a hyper rectangle $\rm{[\bf a,\bf b]} = \{ \rm{\bf z ~|~ \bf a \preceq \bf z \preceq \bf b} \}$ is referred to as a box, when $\rm {\bf 0 \preceq \bf a \preceq \bf b}$.
\end{definition1}

\begin{definition1}\label{def:normal-set}
\emph{(Normal set):} A set $\mathcal{G} \subset \mathcal{R}^N_+$ is normal if for any element $\rm{\bf z} \in \mathcal{G}$, the box $\rm{[\bf 0,\bf z]} \subset \mathcal{G}$.
\end{definition1}

\begin{definition1}\label{def:conormal-set}
\emph{(Conormal set):} A set $\mathcal{H}$ is conormal in $\rm{[\bf 0,\bf b]}$ if and only if the set $\rm{[\bf 0,\bf b]} ~\backslash ~ \mathcal{H}$ is normal.
\end{definition1}

\begin{definition1}\label{def:polyblock}
\emph{(Polyblock):} A set $\mathcal{P} \subset \mathcal{R}^N_+$ is a polyblock if it is the union of all boxes $\rm{[\bf 0,\bf z]}$, $\rm{\bf z} \in \mathcal{V}$, where $\mathcal{V}$ is called the vertex set of $\mathcal{P}$.
\end{definition1}

\begin{definition1}\label{def:projection}
\emph{(Projection):} For any nonempty normal set $\mathcal{G} \subset \mathcal{R}^N_+$ and any vector ${\rm{\bf z}} \in \mathcal{R}^N_+ ~\backslash ~ \mathcal{G}$, $\Phi(\rm{\bf z})$ is called the projection of $\rm{\bf z}$ on $\mathcal{G}$ if $\Phi(\rm{\bf z})$ is on the boundary of $\mathcal{G}$, i.e., $\Phi(\rm{\bf z}) = \lambda \rm{\bf z}$, where $\lambda = \max\{\alpha>0 ~|~ \alpha\rm{\bf z} \in \mathcal{G}\}$.
\end{definition1}

\newtheorem *{definition2}{\hspace{2em}Definition 6}
\begin{definition2}\label{def:monotonic-optimization}
An optimization problem can be categorized into monotonic optimization problems if it can be formulated in the following form:
\begin{equation}\label{equation:standard-monotonic-optimization}
\max \limits_{\rm \bf z} \{ \varphi(\rm \bf z) ~|~ \rm \bf z \in \mathcal{G} \cap \mathcal{H} \},
\end{equation}
where $\varphi({\rm \bf z}): \mathcal{R}^N_+  \rightarrow  \mathcal{R}$ is an increasing function, $\mathcal{G}  \subset  \mathcal{R}^N_+$ is a normal set, and $\mathcal{H}$ is a closed conormal set in $\rm{[\bf 0,\bf b]}$.
\end{definition2}
\begin{figure}[ht]
\centering
\subfigure[]{\includegraphics[width=3.6cm]{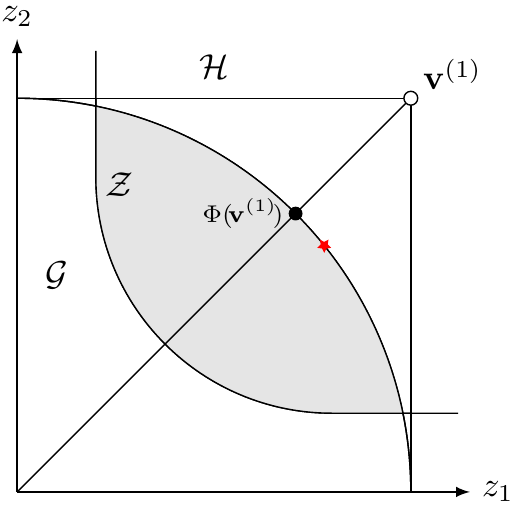}\label{figure:monotonic-figure1}}
\subfigure[]{\includegraphics[width=3.6cm]{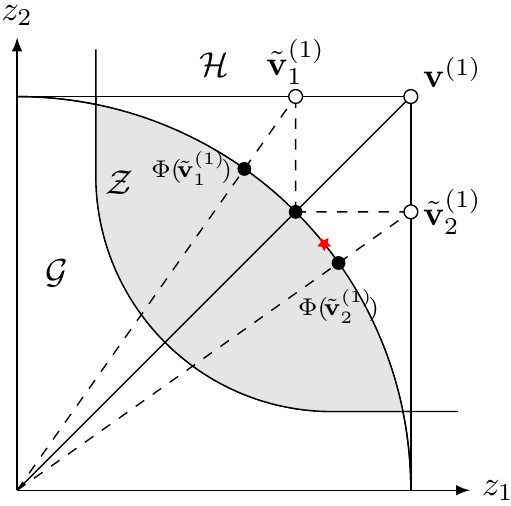}\label{figure:monotonic-figure2}}
\subfigure[]{\includegraphics[width=3.6cm]{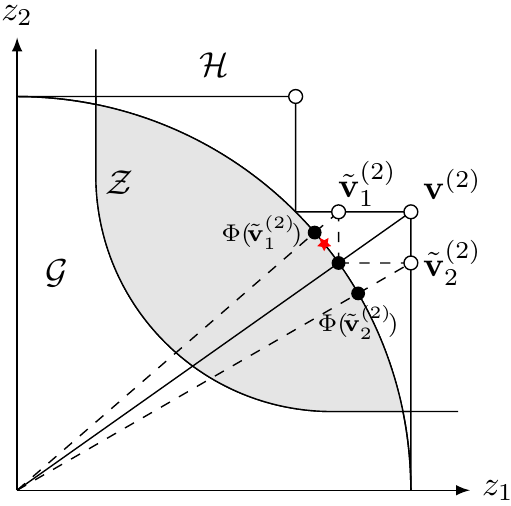}\label{figure:monotonic-figure3}}
\subfigure[]{\includegraphics[width=3.6cm]{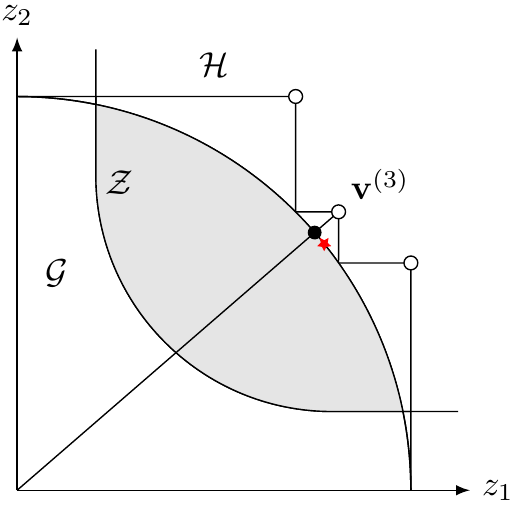}\label{figure:monotonic-figure4}}
\caption{Illustration of the optimal power allocation algorithm based on polyblock outer approximation approach for $N=2$. The red star marks the optimal point on the boundary of the $\mathcal{G}$ in the feasible set $\mathcal{Z}=\mathcal{G}\cap\mathcal{H}$.}\label{figure:monotonic-figure}

\end{figure}

\subsubsection{Optimal Power allocation Algorithm}
Although the objective function of the GLFP problem in (\ref{equation:GLFP-optimization(a)}) is not an monotonic function of ${\bf P}$, there still exists a hidden monotonicity as the function $\psi({\bf x})$ in (\ref{equation:GLFP-optimization(a)}) is monotonically increasing on $\mathcal{R}_+^N$. Nevertheless, the problem in (\ref{equation:GLFP-optimization}) is different from the standard monotonic optimization problem in (\ref{equation:standard-monotonic-optimization}). For the constraints, the feasible set of (\ref{equation:GLFP-optimization(b)})--(\ref{equation:GLFP-optimization(d)}) is not the intersection of a normal set and a conormal set as defined in Definition 6.
\par

To exploit the nature of the hidden monotonicity, where the objective function is an increasing function of the function $\pmb{\bm{\gamma}}(\bm{P})$, we rewrite the problem in (\ref{equation:GLFP-optimization}) in an equivalent form:
\begin{equation}\label{equation:transformed-monotonic-optimization}
\begin{aligned}
\max_{\rm \bf z} \ &\psi({\rm \bf z})\\
\text{s.t.}~\ \; &{\rm \bf z} \in \mathcal{Z},
\end{aligned}
\end{equation}
where $\psi({\rm \bf z})$ is defined in (\ref{equation:phi-function}). The feasible set $\mathcal{Z}$ is given by $\mathcal{Z}=\{{\rm \bf z}~|~ \rm \bf z \in \mathcal{G} \cap \mathcal{H} \}$, where the normal set $\mathcal{G}$ and conormal set $\mathcal{H}$ are spanned by constraints (\ref{equation:GLFP-optimization(b)})--(\ref{equation:GLFP-optimization(d)}) as follows:
\begin{equation*}
\begin{aligned}
\!\mathcal{G}&\!=\!\{{\rm \bf z}| 0 \! \leq \!  z_n \! \leq \! \gamma_n({\bf P}), (\ref{equation:GLFP-optimization(b)}), (\ref{equation:GLFP-optimization(c)}), \gamma_n({\bf P}) \! \leq \! \gamma_n^{max}, \forall n \in \mathcal{N}  \},\\
\!\mathcal{H}&\!=\!\{{\rm \bf z}| z_n \! \geq \! \gamma_n^{min},\forall n \in \mathcal{N} \}.
\end{aligned}
\end{equation*}
\par

With the analysis shown above, we now can design a power allocation algorithm to solve the monotonic optimization problem in (\ref{equation:transformed-monotonic-optimization}) based on the polyblock outer approximation approach \cite{tuy2000Monotonic}. Due to the monotonic property of the objective function, the optimal solution can be attained on the boundary of $\mathcal{G}$ in the feasible set $\mathcal{Z}$ \cite{tuy2000Monotonic}, \cite{zhang2013Monotonic}, which is the basic idea behind the monotonic optimization theory.
\par
\begin{algorithm}
\setlength{\abovedisplayskip}{4pt}
\setlength{\belowdisplayskip}{4pt}
\DontPrintSemicolon
\SetAlgoHangIndent{0pt}
  \caption{Optimal Quality-Driven Power Allocation Algorithm based on Polyblock Outer Approximation}
   \label{alg:Optimal-polyblock}
   \KwIn{$P^{max}, ~\mathcal{P}, ~\mathcal{N}, ~\epsilon$}
   \KwOut{$\bm{P}^*=\{P_n^*\}, ~n\in\mathcal{N}$}
   \textbf{initialization:} Set the iteration index $j=1$. Construct the polyblock $\mathcal{B}^{(1)}$ with vertex set $\mathcal{V}^{(1)}=\{{\bf v}^{(1)}\}$, where the entries of ${\bf v}^{(1)}$ are set as $v_n^{(1)}=\frac{|h_n|^2P^{max}}{\sigma^2}$. \\

   \Repeat{$\frac{||{\bf v}^{(j)}-\Phi({\bf v}^{(j)})||}{||{\bf v}^{(j)}||}\leq \epsilon$}{
    $j = j+1$.\\
    Find $\Phi({\bf v}^{(j-1)})$ based on Algorithm \ref{alg:projection}.\\
    Construct a smaller polyblock $\mathcal{B}^{(j)}$ with vertex set $\mathcal{V}^{(j)}$ by replacing ${\bf v}^{(j-1)}$ in $\mathcal{V}^{(j-1)}$ with $N$ new vertices $\{\tilde{{\bf v}}_1^{(j-1)},\cdots,\tilde{{\bf v}}_N^{(j-1)}\}$, where the $n^{th}$ new vertex is obtained as
    \begin{equation*}
    \tilde{{\bf v}}_n^{(j-1)}\!=\!{\bf v}^{(j-1)}\!+\!(\phi_n({\bf v}^{(j-1)})\!-\!v_n^{(j-1)}){\bf e}_n.
    \end{equation*}\\
    Select ${\bf v}^{(j)}=\arg \max \ \{\psi(\Phi({\bf v}))|{\bf v}\in\mathcal{V}^{(j)}\}$.\\

    }
   ${\bf v}^{*}=\Phi({\bf v}^{(j)})$ and compute the optimal power allocation vector ${\bf P}^*$\ by solving $\phi_n({\bf v}^{(j)})=\frac{f_n({\bf P}^*)}{\xi_n({\bf P}^*)}$.
\end{algorithm}

Therefore, the key to this algorithm is finding the boundary of $\mathcal{G}$. Here, we attempt to handle it by constructing a sequence of polyblocks. Specifically, a polyblock $\mathcal{B}^{(1)}$, which includes only one vertex ${\bf v}^{(1)}$ with positive entries in the vertex set $\mathcal{V}^{(1)}$, is initially constructed to enclose the feasible set $\mathcal{Z}$. According to Definition 5, the projection of ${\bf v}^{(1)}$ on the boundary of $\mathcal{G}$ can be found, denoted as $\Phi({\bf v}^{(1)})$. Then a smaller polyblock $\mathcal{B}^{(2)}$ is constructed based on $\mathcal{B}^{(1)}$, by replacing ${\bf v}^{(1)}$ with $N$ new vertices $\tilde{\mathcal{V}}^{(1)}=\{\tilde{{\bf v}}_1^{(1)},\cdots,\tilde{{\bf v}}_N^{(1)}\}$. Thus, the vertex set $\mathcal{V}^{(2)}$ of $\mathcal{B}^{(2)}$ is $\mathcal{V}^{(2)}=(\mathcal{V}^{(1)}\backslash{\bf v}^{(1)}) \cup \tilde{\mathcal{V}}^{(1)}$. Note that the new vertex $\tilde{{\bf v}}_n^{(1)}$ is generated by replacing the $n^{th}$ entry of ${\bf v}^{(1)}$ with the $n^{th}$ entry of $\Phi({\bf v}^{(1)})$, which can be expressed as:
\begin{equation*}
\tilde{{\bf v}}_n^{(1)}={\bf v}^{(1)}+(\phi_n({\bf v}^{(1)})-v_n^{(1)}){\bf e}_n,
\end{equation*}
where $\phi_n({\bf v}^{(1)})$ is the $n^{th}$ entry of $\Phi({\bf v}^{(1)})$, and ${\bf e}_n$ is the $n^{th}$ unit vector of $\mathcal{R}^n$ with a non-zero entry only at index $n$. After these steps, the smaller polyblock $\mathcal{B}^{(2)}$ will still enclose the feasible set $\mathcal{Z}$. Then, the optimal vertex, whose projection maximizes the objection function in (\ref{equation:transformed-monotonic-optimization}), is selected from $\mathcal{V}^{(2)}$, i.e., ${\bf v}^{(2)}=\arg \max \ \{\psi(\Phi(\bf v))|{\bf v}\in\mathcal{V}^{(2)}\}$. Repeating this procedure, we can construct a sequence of polyblocks gradually outer approximating the feasible set:
\begin{equation*}
\mathcal{B}^{(1)}\supset\mathcal{B}^{(2)}\supset\cdots\supset\mathcal{B}^{(j)}\supset\cdots\supset\mathcal{Z}.
\end{equation*}
The algorithm terminates when $\frac{||{\bf v}^{(j)}-\Phi({\bf v}^{(j)})||}{||{\bf v}^{(j)}||}\leq \epsilon$, where $\epsilon \geq 0$ is the error tolerance.
\par

To better understand the algorithm, we illustrate it for a case of $N=2$ in Fig. \ref{figure:monotonic-figure}. The complete algorithm is presented in Algorithm \ref{alg:Optimal-polyblock}. In the initialization phase, we need to find a box (polyblock) $\rm{[\bf 0,{\bf v}^{(1)}]}$ to contain the feasible set $\mathcal{Z}$. The simple way we adopt is to set the $n^{th}$ entry ${v}_n^{(1)}$ be the upper bound of the achievable SINR of the $n^{th}$ NOMA UE, i.e., $v_n^{(1)}=\max\limits_{{\bf P}}\gamma_n({\bf P})=\frac{|h_n|^2P^{max}}{\sigma^2}$. Other tighter initialization box can be set at the cost of additional computation. The convergence of Algorithm 1 is guaranteed according to \cite{tuy2000Monotonic}, \cite{zhang2013Monotonic}.
\par

\begin{algorithm}
\setlength{\abovedisplayskip}{4pt}
\setlength{\belowdisplayskip}{4pt}
  \caption{Projection Algorithm}
  \DontPrintSemicolon
  \SetAlgoHangIndent{0pt}
   \label{alg:projection}
   \KwIn{${\bf v}^{(j)}, ~P^{max}, ~\mathcal{P}, ~\mathcal{N}, ~\delta$}
   \KwOut{$\Phi({\bf v}^{(j)})$}
   \textbf{initialization:} Set the iteration index $i=0$. Set $\lambda_j^0 = 0$.\\
   \Repeat{$\min\limits_{1\leq n \leq N}\{f_n({\bf P}^{i-1})-\lambda_j^{i-1}{v}_n^{(j)}\xi_n({\bf P}^{i-1})\}\leq \delta$}{

    ${\bf P}^{i}=\mathop{\arg\max}\limits_{{\bf P} \in \mathcal{P}} \{ \min\limits_{1\leq n \leq N}\{f_n({\bf P})-\lambda_j^{i}{v}_n^{(j)}\xi_n({\bf P})\} \}$.\\

    $\lambda_j^{i+1}=\min\limits_{1\leq n \leq N}\frac{f_n({\bf P}^{i})}{{v}_n^{(j)}\xi_n({\bf P}^{i})}$.\\

    $i=i+1$.\\
    }
   The projection is $\Phi({\bf v}^{(j)})=\lambda_j^{i-1}{\bf v}^{(j)}$.
\end{algorithm}
The projection $\Phi({\bf v}^{(j)})$ needs to be calculated, in the process of constructing smaller polyblocks and judging the termination criterion at step 5 and 6, respectively. This is not straightforward as the boundary of $\mathcal{G}$ is not explicitly known. However, recalling the Definition 5, $\Phi({\bf v}^{(j)})=\lambda_j{\bf v}^{(j)}$ can be obtained by solving
\begin{equation}
\begin{aligned}
\lambda_j   &= \max\{\alpha>0 ~|~ \alpha {\bf v}^{(j)} \in \mathcal{G}\}\\
            &= \max\{\alpha>0 ~|~ \alpha \leq \min_{1\leq n \leq N}\frac{f_n({\bf P})}{{v}_n^{(j)}\xi_n({\bf P})},~ \forall {\bf P} \in \mathcal{P}\}\\
            &= \max_{{\bf P} \in \mathcal{P}} \min_{1\leq n \leq N}\frac{f_n({\bf P})}{{v}_n^{(j)}\xi_n({\bf P})},
\end{aligned}
\end{equation}
where $\mathcal{P}$ is characterized by system power constraints and SINR constraints ($\pmb{\bm{\gamma}}^{min}, \pmb{\bm{\gamma}}^{max}$). This problem can be solved through the Dinkelbach algorithm in \cite{dinkelbach1967on}. The details are summarized in Algorithm \ref{alg:projection}.
\par

By exploiting the monotonic property of the problem in (\ref{equation:GLFP-optimization}), Algorithm \ref{alg:Optimal-polyblock} can find the global optimal solution with drastically reduced complexity. However, its computational complexity still increases exponentially with the dimension of the problem $N$. This is another factor to be considered when we adopt NOMA users grouping strategies. To approach the solution of the problem in (\ref{equation:GLFP-optimization}) with polynomial time computational complexity, we design a suboptimal fast algorithm based on a greedy strategy in the following.
\begin{algorithm}
  \caption{Fast Greedy Algorithm for Power Allocation}
  \DontPrintSemicolon
  \SetAlgoHangIndent{0pt}
   \label{alg:suboptimal-greedy}
   \KwIn{$P^{max}, ~{\bm{Q}}^{min},~{\bm{Q}}^{max}, ~\mathcal{N}, ~p_l=P^{max}/L$}
   \KwOut{$\bm{P}^*=\{P_n^*\}, ~n\in\mathcal{N}$}
   \textbf{initialization:}$~P_n^*=0, ~n_d = |\mathcal{N}|$.\\
   \For {$n = 1 : |\mathcal{N}|$ }{
    \Repeat{$Q_{n_d}(\bm{P}^*) \geq Q_{n_d}^{min} ~{\rm or}~ P^{max}=0$}{
        $P_{n_d}^*=P_{n_d}^*+p_l$.\\
        $P^{max} = P^{max}-p_l$.\\}
        \If{$Q_{n_d}(\bm{P}^*) < Q_{n_d}^{min} ~{\rm and}~ P^{max}=0$}{
        report infeasibility, {\bf break}.\\
        }
    $n_d = n_d-1$.\\
   }
   \Repeat{$P^{max}=0$}{
    \For {$n = 1 : |\mathcal{ N}|$ }{
        $\bm{P}=\bm{P}^*, ~{\rm except} ~P_n=P_n^*+p_l $.\\
            \eIf {${\bm{Q}}^{min} \leq {\bm{Q}}(\bm{P}) \leq {\bm{Q}}^{max}$}{
                Calculate $Q(\bm{P})$ using (\ref{equation:original-optimization-a}).\\
                }{
                $Q(\bm{P})=-\infty$.\\
            }
        }
    $n_d = \arg \max_{n \in \mathcal{N}} Q(\bm{P})$.\\
    $P_{n_d}^*=P_{n_d}^*+p_l$.\\
    $P^{max} = P^{max}-p_l$.\\
   }
\end{algorithm}
\subsection{Fast suboptimal solution}
\subsubsection{Fast Greedy Algorithm for Power allocation}
To reduce complexity of the problem in (\ref{equation:GLFP-optimization}), we can discretize the power first, namely the total power $P^{max}$ is equally split into $L$ power blocks (PBs) with value $p_l=P^{max}/L$. In fact, the discrete power settings are common in the literatures \cite{sokun2017optimization} and in practice (e.g., in 3GPP LTE).
\par

Now we can design a fast algorithm to allocate the discrete PBs among NOMA UEs in two phases. In phase \uppercase\expandafter{\romannumeral1}, we allocate the PBs to the $N^{th}$ UE until its quality requirement $Q_N^{min}$
is met. Then we allocate the PBs to the ${N-1}^{th}$ UE until $Q_{N-1}^{min}$ is satisfied. This procedure is repeated until the $1^{th}$ UE's quality requirement is satisfied. The order for power allocation is determined according to the property of NOMA systems. With SIC, the UE's capacity would not be interfered by the UE with a worse channel condition. In phase \uppercase\expandafter{\romannumeral2}, the remaining PBs are allocated in a greedy manner to maximize the average PSNR of all UEs. The power budget and quality requirements should not be violated in these two phases. The proposed suboptimal algorithm is summarized in Algorithm \ref{alg:suboptimal-greedy}.

\subsubsection{Computational Complexity Analysis}
In Algorithm 3, steps 2--11 require complexity of $\mathcal{O}(NL)$, and steps 12--24 require complexity of $\mathcal{O}((N^2+N)L)$. Hence, computational complexity of the proposed fast algorithm is $\mathcal{O}(N^2L+2NL)$.

\section{Performance Evaluation}\label{section:performance-evaluation}
We carry out simulations to evaluate performance of the proposed scheme, which is based on the cross-layer design and power allocation algorithms, under various scenarios. We consider a single-cell NOMA system with a total of six UEs, that are equally partitioned into two zones, where UEs are randomly and uniformly distributed. Thus, we have $M=3$ NOMA groups, each of which has $N=2$ NOMA UEs. The bandwidth and average power budget of each NOMA group are $B =140$ kHz and $P=1$ W, respectively. The path-loss exponent is set as $\eta=2$. As discussed in \ref{subsection:NOMA Wireless Network with User Grouping}, the AWGN for each UE has the same variance $\sigma^2$, and the channel signal-to-noise ratio (SNR) is defined as $10\,\log_{10}(P/\sigma^2)$. The number of PBs in Algorithm 3 is set as $L=100$. The AMC scheme at the PHY layer is characterized by a {\em rate adjustment} $c_1=0.905$ and a {\em SNR gap} $c_2=1.34$ \cite{mazzotti2012multiuser}.
\par
We encode six video sequences, one for each UE, with common intermediate format (CIF) resolution and the frame-rate of 30 fps (frame per second). Similar to \cite{lin2017cross}, we select three sequences with lower spatial-temporal content complexity (i.e., `Foreman', `Ice' and `Crew'), as well as three sequences with higher spatial-temporal content complexity (i.e., `Football', `Mobile' and `Soccer'). Each sequence is encoded by the joint scalable video model (JSVM) software \cite{JSVM} and the GOP size is set to 8. The encoded stream contains one base layer and three enhancement layers, with quantization parameters of 40, 34, 28 and 22, respectively. Each enhancement layer is further split into five MGS layers with the vector [4 3 2 3 4]. Other parameters for the cross-layer framework are set as follows. The number of bytes per RS codeword in the UXP scheme is set to maximum as 255. Moreover, we set the size of an RTP packet to 1400 bytes and the RTP packet loss rate $p_{rtp}=5\%$. PSNR is calculated using (\ref{equation:PSNR}) with the average luminance MSE.
\par
For these simulations, we set the quality constraints ${\bf Q}^{min}$ and ${\bf Q}^{max}$ in (\ref{equation:original-optimization}), as PSNR of decoding the base layer and all layers, respectively. Note that the optimal rate ${F}_n^*({\bf P}^*)$ may not be achievable, as the SVC scheme only supports a discrete set of rate values. In practice \cite{lin2017cross}, the largest achievable rate in the discrete set, which is smaller than ${F}_n^*({\bf P}^*)$, is selected.
%
\begin{table}[htb]
\vspace{-0.6cm}
\centering
\caption{Power allocation coefficient of the weaker UE}\label{tab:power-allocation}
\vspace{-0.1cm}
\begin{tabular}{|c|C{0.7cm}|C{0.7cm}|C{0.7cm}|C{0.6cm}|C{0.6cm}|C{0.6cm}|}
     \hline

     \multirow{2}*{{\diagbox{Group}{SNR}}} & \multicolumn{3}{c|}{Proposed-POA} & \multicolumn{3}{c|}{Proposed-Greedy}\\
     \cline{2-7}  & 10dB & 20dB & 30dB & 10dB & 20dB & 30dB\\
     \hline
     1 & 0.584 & 0.779 & 0.898 & 0.580 & 0.770 & 0.890 \\
     \hline
     2 & 0.579 & 0.768 & 0.880 & 0.570 & 0.760 & 0.880 \\
     \hline
     3 & 0.629 & 0.835 & 0.940 & 0.620 & 0.830 & 0.930 \\
     \hline
\end{tabular}
\vspace{-0.2cm}
\end{table}

\subsection{Analysis of Power Allocation Results}\label{subsection:power-allocation}
We will first evaluate the power allocation result in the proposed scheme based on polyblock outer approximation (Proposed-POA), and the one based on the greedy strategy (Proposed-Greedy). Table \ref{tab:power-allocation} provides the power allocation coefficient of the weaker UE, which illustrates several insights. First, due to the size of NOMA group ($N=2$) under
consideration and the use of discrete PBs ($L=100$), the power allocation
result achieved by the suboptimal algorithm is similar to that of the optimal algorithm. Second, recall that in OMA schemes, conventional power allocation strategies (e.g., water filling strategies) allocate more power to better UEs. However, in the proposed scheme, the weaker UE receives more transmission power, which ensures that the weaker UE can decode its streaming information directly by treating information of other UEs as noise. Moreover, both algorithms allocate more power to the weaker UE when SNR increases. Based on the expression of SINR in (\ref{equation:SINR}), this result indicates that MUI is much more intense at high SNR that the power allocation coefficient of the better UE has to be reduced.
\begin{figure}[!h]
\centering
\includegraphics[width=6.3cm]{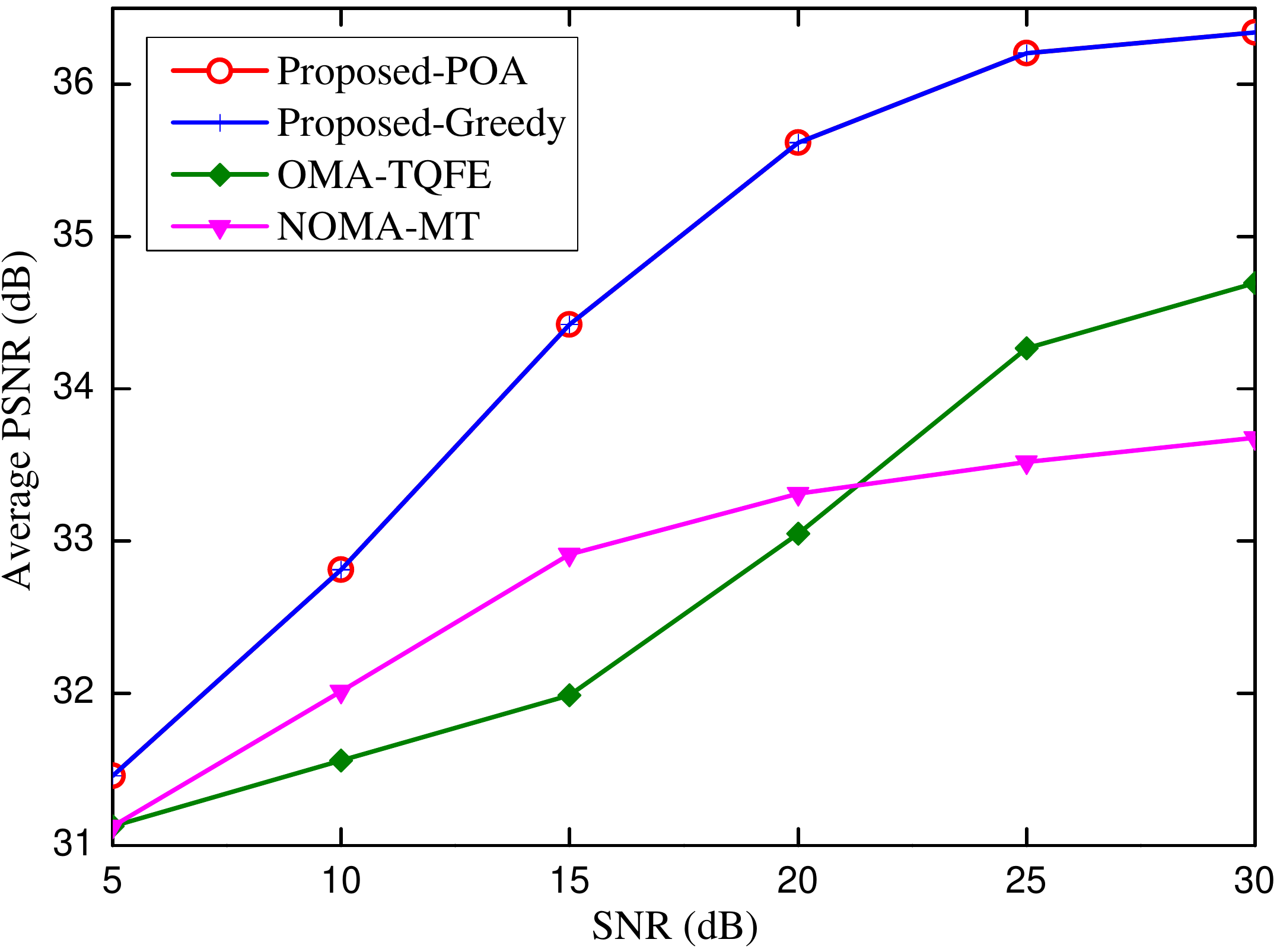}
\vspace{-0.1cm}
\caption{Average PSNR performance comparison among different schemes.}\label{figure:system-compare}
\vspace{-0.15cm}
\end{figure}

\begin{figure}[htb]
\centering
\includegraphics[width=6.3cm]{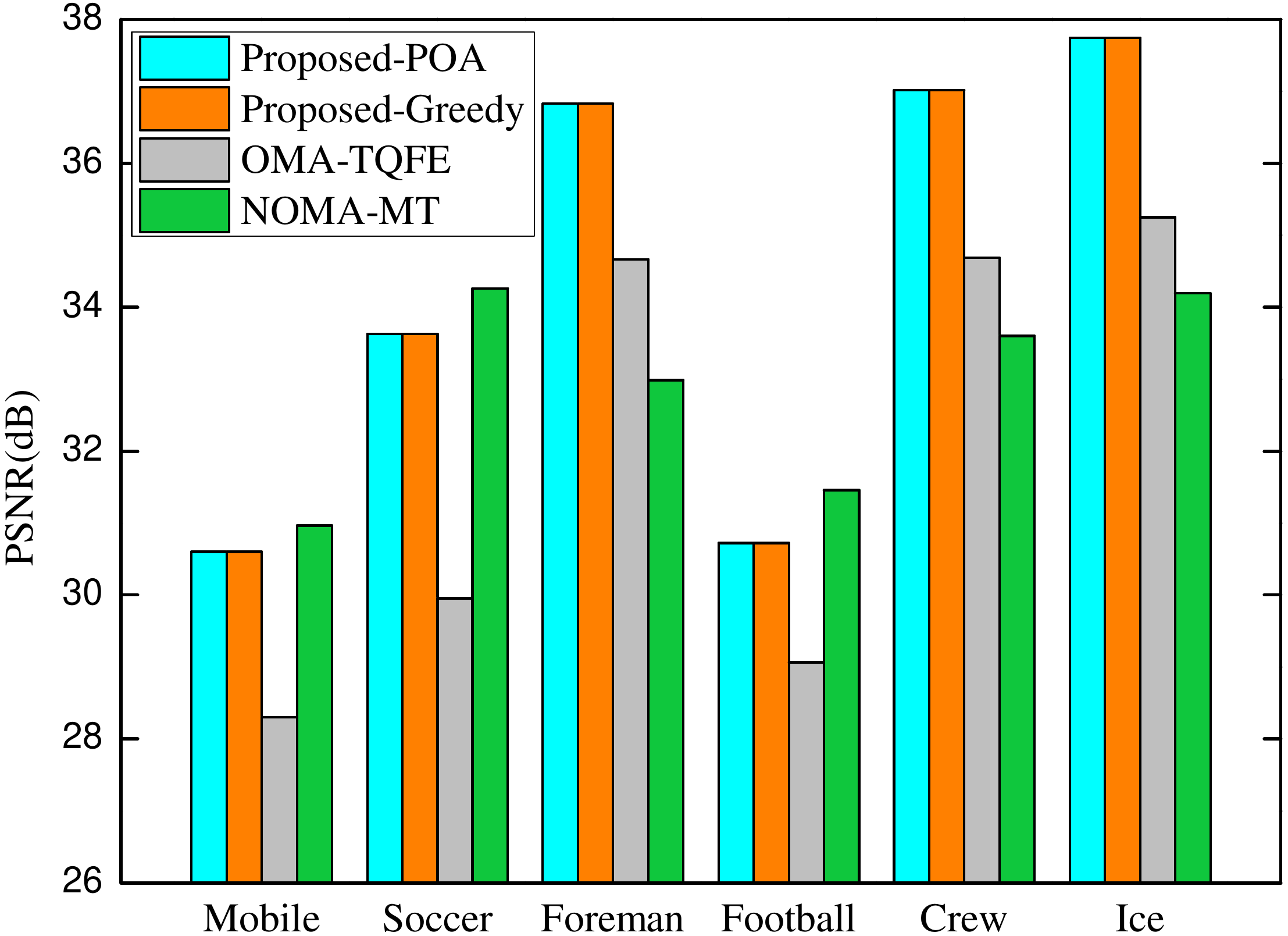}
\vspace{-0.1cm}
\caption{PSNR of each video sequence for different schemes, SNR\,=\,15dB.}\label{figure:peruser-performance}
\end{figure}

\subsection{Performance Comparison}\label{subsection:system-comparison}

We compare performance of the proposed two algorithms against two reference schemes. One is an OMA scheme in which the resources are allocated for tradeoff between quality fairness and efficiency (OMA-TQFE) \cite{lin2017cross}. The tradeoff is controlled by a parameter $\varpi$. For comparison, we consider the extreme case $\varpi=\infty$, which implies that the scheme maximizes the system average PSNR. Another is an NOMA scheme
designed for maximizing system throughput for data transmission (NOMA-MT) \cite{wang2016power}, in which power is allocated to the weak UE for merely guaranteeing its minimum QoS requirement. Note that the design of NOMA-MT cannot be straightly applied into video transmission. For fair comparison, we implement NOMA-MT with our proposed cross-layer video delivery design.
\par

In these simulations, we consider that weaker UEs request `Foreman', `Ice' and `Crew', and better UEs request the remaining sequences. System performance is evaluated by averaging PSNR over all received video sequences, as Fig. \ref{figure:system-compare} shows. Note that performance of Proposed-Greedy approaches that of Proposed-POA, which is consistent with  power allocation results in Table \ref{tab:power-allocation}. Over the entire range of SNR, the proposed schemes outperform OMA-TQFE and NOMA-MT between 0.5$\sim$2.6\,dB in terms of the average PSNR. When SNR is low at 5\,dB, all schemes have similar performance. This is due to the poor channel condition, which can only support transmission of one base layer and very few MGS layers. At medium SNR, OMA-TQFE is inferior to NOMA-MT as NOMA can achieve higher spectral efficiency than OMA. As the channel capacity improves towards high SNR, OMA-TQFE outperforms NOMA-MT, since the former is inherently designed for maximizing video quality, while the latter is designed for maximizing throughput.
\par

\begin{figure}[htb]
\centering
\subfigure{\includegraphics[width=4.3cm]{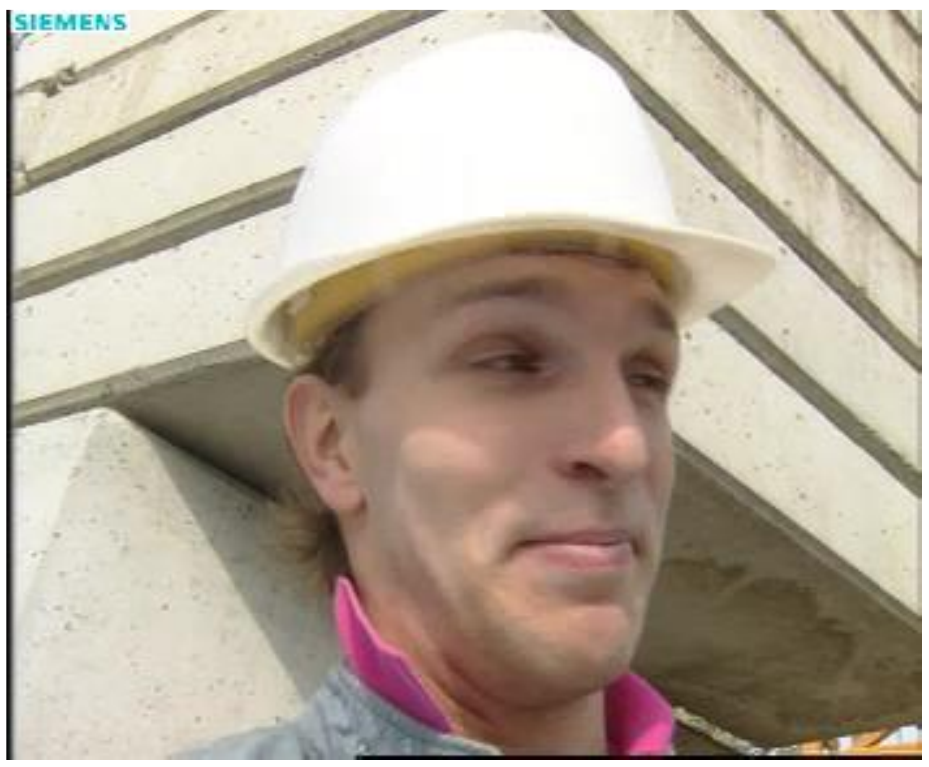}\label{figure:single_fix_rate_SDR_a}}
\subfigure{\includegraphics[width=4.3cm]{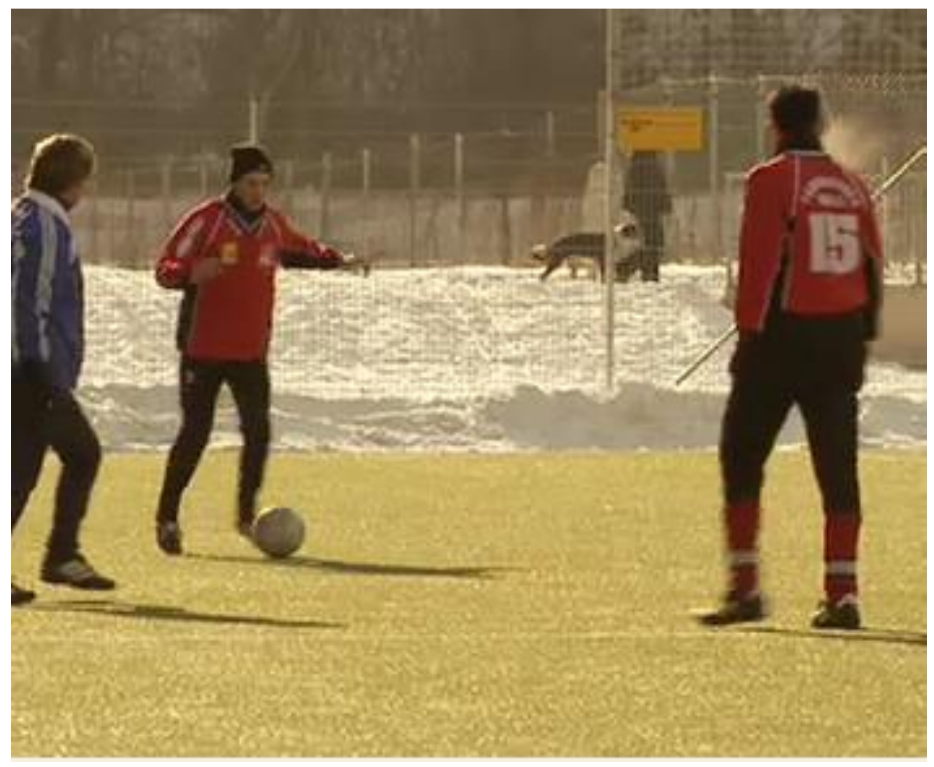}\label{figure:single_fix_rate_SDR_b}}
\subfigure{\includegraphics[width=4.3cm]{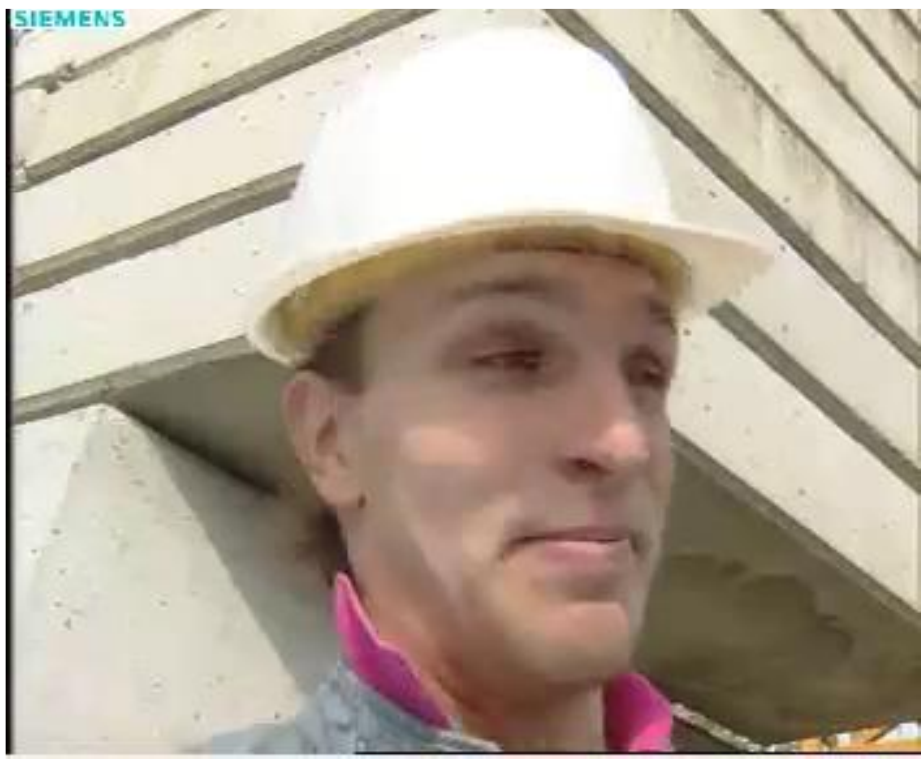}\label{figure:single_fix_rate_SDR_a}}
\subfigure{\includegraphics[width=4.3cm]{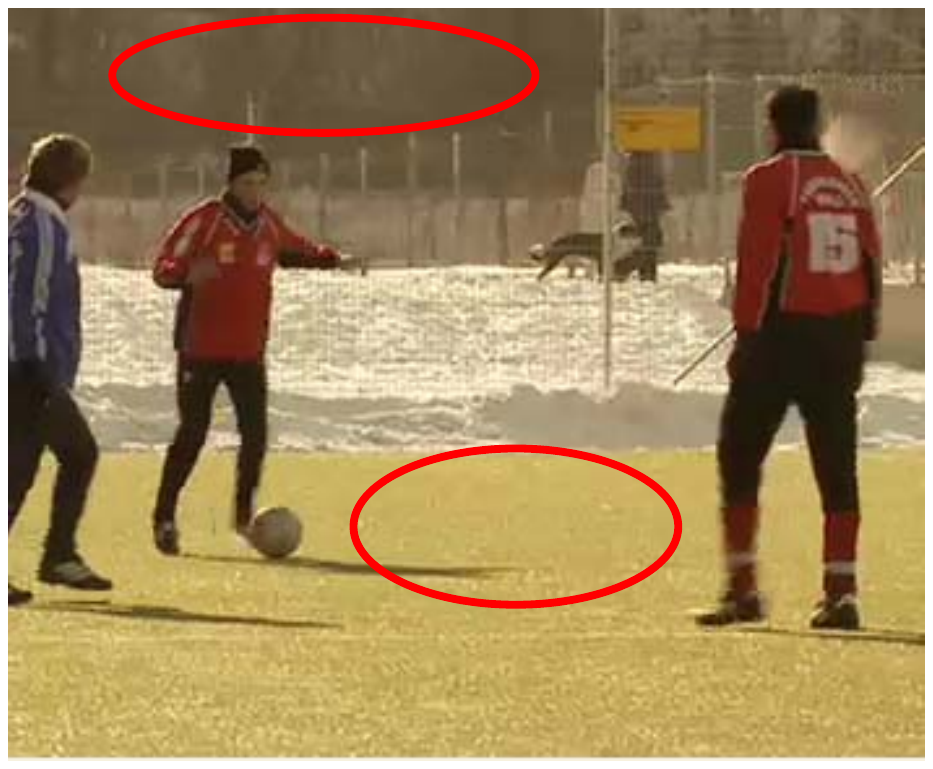}\label{figure:single_fix_rate_SDR_b}}
\subfigure{\includegraphics[width=4.3cm]{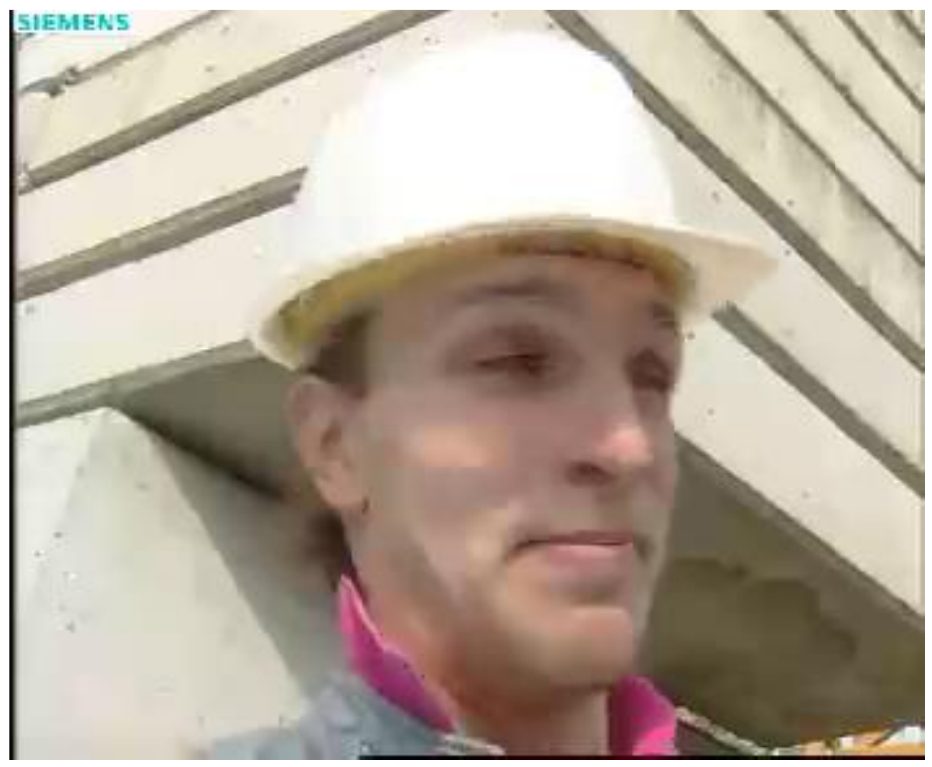}\label{figure:single_fix_rate_SDR_a}}
\subfigure{\includegraphics[width=4.3cm]{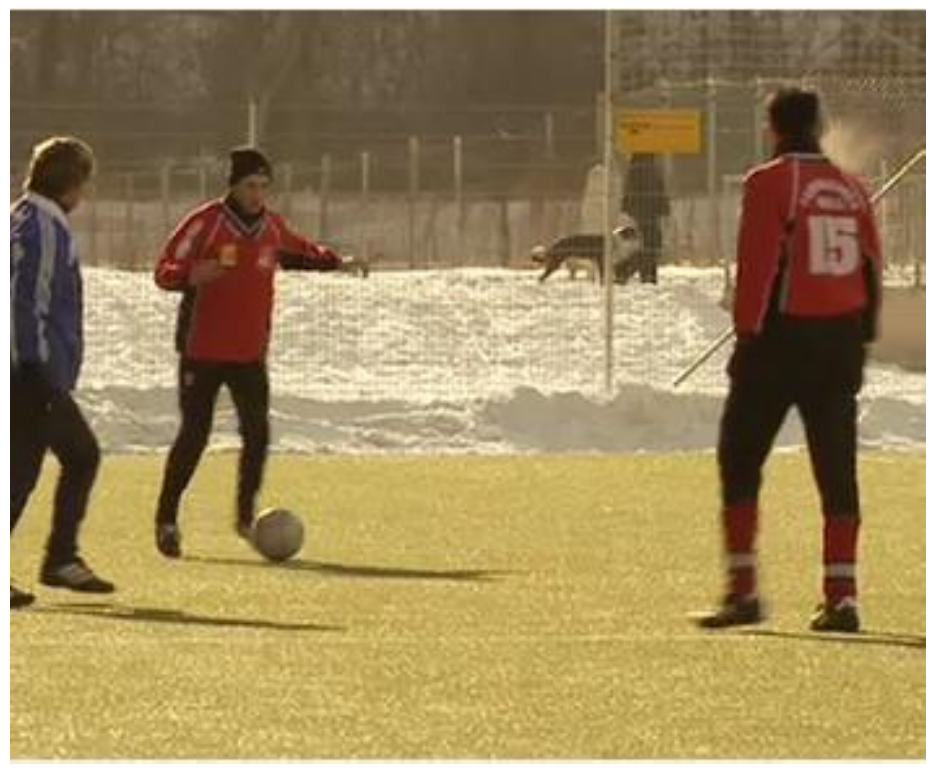}\label{figure:single_fix_rate_SDR_b}}
\caption{Visual comparison among the proposed scheme (first row), OMA-TQFE (second row) and NOMA-MT (third row), for `Foreman' (left) and `Soccer' (right), SNR\,=\,15dB.}\label{figure:visual-comparison}
\end{figure}

Fig. \ref{figure:peruser-performance} illustrates PSNR of each video sequence corresponding to various schemes, at SNR\,=\,15\,dB. The proposed schemes outperform OMA-TQFE for all video sequences. Although PSNR values of NOMA-MT are similar with those of the proposed schemes for sequences requested by better UEs, performance of NOMA-MT drastically degrades for the sequences requested by weaker UEs. This is due to the power allocation strategy in NOMA-MT that sacrifices throughput of weaker UEs to enhance throughput of better UEs, which is unfair to weaker UEs.
\par
Fig. \ref{figure:visual-comparison} shows reconstructed frames of `Foreman' and `Soccer', requested by weaker and better UEs, respectively. We select frames obtained by the greedy algorithm to represent the result of the proposed scheme. It is apparent that the proposed scheme improves received quality for `Foreman'. For `Soccer', the frame reconstructed by OMA-TQFE contains blurred regions (the lawn far away and trees in the background). Although NOMA-MT ensures satisfactory quality similarly with the proposed scheme for `Soccer', it reconstructs the `Foreman' frame with poor quality.
\vspace{-0.1cm}
\subsection{Impacts of User Grouping Strategies}\label{subsection:impacts of user grouping}
We also investigate impacts of user grouping strategies, which are based on requested video content characteristics, on performance of the proposed NOMA system. Three cases are considered: 1) weaker UEs request sequences
with lower spatial-temporal content complexity and better UEs request sequences with higher content complexity (WLBH); 2) weaker UEs request sequences with higher content complexity and better UEs request sequences with lower content complexity (WHBL); 3) weaker UEs and better UEs request sequences randomly (WRBR). Table \ref{tab:user-grouping} presents PSNR results in these three cases, obtained by Proposed-Greedy.
\par

An important observation is that WLBH outperforms other two cases and WHBL achieves the worst performance. This is due to the fact that sequences with higher spatial-temporal content complexity need to be encoded with higher rate. Thus, performance would improve only if those sequences are transmitted through channels with better channel conditions.
\par

It is worth noting that this phenomenon provides the guidelines for grouping NOMA UEs dynamically, especially for scenarios with high mobility, where locations of UEs are not confined to specific regions. The above observations indicate that it is more beneficial to dynamically group weaker UEs requesting sequences of lower spatial-temporal content complexity, together with better UEs requesting sequences of higher content complexity.
\begin{table}[htb]
\vspace{-0.2cm}
\centering
\caption{PSNR performance under different user grouping strategies.}\label{tab:user-grouping}
\begin{tabular}{|c|C{0.7cm}|C{0.7cm}|C{0.7cm}|C{0.6cm}|C{0.6cm}|C{0.6cm}|}
     \hline

     \multirow{2}*{{\diagbox{Video}{SNR}}} & \multicolumn{2}{c|}{WLBH} & \multicolumn{2}{c|}{WRBR} & \multicolumn{2}{c|}{WHBL}\\
     \cline{2-7}  & 15dB & 25dB & 15dB & 25dB & 15dB & 25dB\\
     \hline
     Mobile & 30.54 & 32.03  & 30.10 & 32.01 & 29.66 & 31.98\\
     \hline
     Football & 30.74 & 33.27  & 30.71 & 33.18 & 30.59 & 33.12\\
     \hline
     Soccer & 33.59 & 34.52  & 31.88 & 34.35 & 31.88 & 35.49\\
     \hline
     Crew & 37.00 & 39.05 & 34.15 & 36.99 & 34.68 & 36.83 \\
     \hline
     Foreman & 36.78 & 38.54 & 36.70 & 38.44 & 36.54 & 38.37 \\
     \hline
     Ice & 37.73 & 39.48 & 36.99 & 39.35 & 34.17 & 37.61 \\
     \hline
     \bf{ Average} & \bf 34.40 & \bf 36.15 & \bf 33.42 & \bf 35.72 & \bf 32.92 & \bf 35.57  \\
     \hline
\end{tabular}
\vspace{-0.05cm}
\end{table}
%

\section{Conclusion}\label{section:conclusion}
This paper has presented the design of a quality-driven scalable video transmission scheme over the multi-user NOMA system. To enable practical NOMA systems, a user grouping strategy has been proposed to reduce implementation complexity. Integrating the characteristics of scalable video streams with the SC in NOMA, a cross-layer power allocation optimization problem has been formulated as a quality maximization problem measured by PSNR. Based on the hidden monotonic property of the non-concave problem, a global optimal algorithm with polyblock outer approximation has been developed. Furthermore, a suboptimal greedy algorithm has also been implemented to approach the solution with polynomial time complexity. Simulation results have shown great performance improvements over existing schemes. The insights, that we derived from the simulation results, have provided useful guidance for designing dynamic user grouping strategies based on the video content.
\par
In the future, we plan to study the application of multi-input and multi-output (MIMO) technology to our NOMA-based video transmission system.

\section*{Acknowledgements}
This work was supported in part by the National Science Foundation of China (No.91538203, No.61390513, No.61771445) and the Fundamental Research Funds for the Central Universities.


%

%
%
%
%
%

\ifCLASSOPTIONcaptionsoff
  \newpage
\fi


\footnotesize
\begin{thebibliography}{99}

%
\bibitem{white2014rethink}
White Paper, ¡°Rethink Mobile Communications for 2020+,¡± {\em Future Mobile Communication Forum 5G SIG}, Nov. 2014. Available: \url{http://www.future-forum.org/dl/141106/whitepaper.zip}.

\bibitem{Higuchi2015non}
K. Higuchi and A. Benjebbour, ``Non-orthogonal multiple access (NOMA) with successive interference cancellation for future radio access,'' {\em IEEE Trans. Commun.}, vol. 98, no. 3, pp. 403--414, March. 2015.

\bibitem{islam2017power}
S. M. R. Islam, N. Avazov, O. A. Dobre, and K.-S. Kwak, ``Power domain
non-orthogonal multiple access (NOMA) in 5G systems: Potentails
and challenges,'' {\em IEEE Commun. Surveys Tuts.}, vol. 19, no. 2, pp. 721--742, May. 2017.

\bibitem{caire2003on}
G. Caire and S. Shamai, ``On the achievable throughput of a multiantenna Gaussian broadcast channel,'' {\em IEEE Trans. Inf. Theory}, vol. 49, no. 7, pp. 1692--1706, July. 2003.



\bibitem{wang2016power}
C. L. Wang, J. Y. Chen, and Y. J. Chen, ``Power allocation for a downlink non-orthogonal multiple access system,'' {\em IEEE Wireless Commun. Lett.}, vol. 5, no. 5, pp. 532--535, Oct. 2016.

\bibitem{ding2016impact}
Z. Ding, P. Fan, and H. V. Poor, ``Impact of user pairing on 5G nonorthogonal multiple access downlink transmissions,'' {\em IEEE Trans. Veh. Technol.}, vol. 65, no. 8, pp. 6010--6023, Aug. 2016.


\bibitem{lv2016cooperative}
L. Lv, J. Chen, and Q. Ni, ``Cooperative non-orthogonal multiple access in cognitive radio,'' {\em IEEE Commun. Lett.}, vol. 20, no. 10, pp. 2059--2062, Oct. 2016.


\bibitem{elbamby}
M. S. ElBamby, M. Bennis, W. Saad, M. Debbah, and M. Latva-aho, ``Resource optimization and power allocation in full duplex non-orthogonal multiple access (FD-NOMA) networks,'' {\em IEEE J. Sel. Areas Commun.}, vol. PP, no. 99, pp. 1--1, 2017.

\bibitem{inc2016cisco}
Cisco, Visual Networking Index, ``Forecast and Methodology, 2015-2020,'' June. 2016.

\bibitem{schierl2005wireless}
T. Schierl, H. Schwarz, D. Marpe, and T. Wiegand, ``Wireless Broadcasting Using The Scalable Extension of H.264/AVC,''  in {\em Proc. IEEE Int. Conf. Multimedia and Expo}, July. 2005. pp. 884--887.

\bibitem{schwarz2007overview}
H. Schwarz, D. Marpe and T. Wiegand, ``Overview of the scalable video
coding extension of the H.264/AVC standard,'' {\em IEEE Trans. Circuits Syst. Video Technol.}, vol. 17, no. 9, pp. 1103--1120, Sep. 2007.

\bibitem{JSVM}
SVC Reference Software, [online], Available: \url{https://www.hhi.fraunhofer.de/en/departments/vca/research-groups/image-video-coding/research-topics/svc-extension-of-h264avc/jsvm-reference-software.html}.

\bibitem{vqeg2003final}
Video Quality Experts Group. ``Final Report from the Video Quality Experts Group on the Validation of Objective Models of Video Quality Assessment Phase II,'' Aug. 2003.

\bibitem{stuhlmuller2000analysis}
K. Stuhlmuller, N. Farber, M. Link, and B. Girod, ``Analysis of video transmission over lossy channels,'' {\em IEEE J. Selec. Areas Commun.}, vol. 18, no. 6, pp. 1012--1032, Jun. 2000.

\bibitem{mansour2008channel}
H. Mansour, V. Krishnamurthy, P. Nasiopoulos, ``Channel aware multi-user scalable video streaming over lossy under-provisioned channels: modeling and analysis,'' {\em IEEE Trans. Multimedia}, vol. 10, no. 7, pp. 1366--1381, Nov. 2008.

\bibitem{he2002joint}
Z. He, J. Cai, and C. W. Chen, ``Joint source channel rate-distortion analysis for adaptive mode selection and rate control in wireless video coding,'' {\em IEEE Trans. Circuits Syst. Video Technol.}, vol. 12, no. 6, pp. 511--523, Jun. 2002.

\bibitem{mazzotti2012multiuser}
M. Mazzotti, S. Moretti, and M. Chiani, ``Multiuser resource allocation with adaptive modulation and LDPC coding for heterogeneous traffic in OFDMA downlink,'' {\em IEEE Trans. Commun.}, vol. 60, no. 10, pp. 2915--2925, Oct. 2012.

\bibitem{tuy2000Monotonic}
H. Tuy, ``Monotonic optimization: Problems and solution approaches,'' {\em SIAM J. Optim.}, vol. 11, no. 2, pp. 464--494, 2000.

\bibitem{zhang2013Monotonic}
Y. J. A. Zhang, L. Qian, and J. Huang, ``Monotonic optimization in communication and networking systems,'' {\em Found Trends Netw.}, vol. 7, no. 1, pp. 1--75, Oct. 2013.

\bibitem{dinkelbach1967on}
W. Dinkelbach, ``On nonlinear fractional programming,'' {\em Management Science}, vol. 13, pp. 492--498, Mar. 1967.

\bibitem{sokun2017optimization}
H. Sokun, E. Bedeer, R. Gohary and H. Yanikomeroglu, ``Optimization of Discrete Power and Resource Block Allocation for Achieving Maximum Energy Efficiency in OFDMA Networks'', {\em IEEE Access}, vol. 5, pp. 8648--8658, May. 2017.

\bibitem{lin2017cross}
K. Lin and S. Dumitrescu, ``Cross-layer Resource Allocation for Scalable Video over OFDMA Wireless Networks: Trade-off between Quality Fairness and Efficiency.'' {\em IEEE Trans. Multimedia}, vol. 19, no. 7, pp. 1654--1669, July. 2017.




\end{thebibliography}
\end{document}